\documentclass[aps,pra,a4paper,twocolumn,amsmath,amssymb,showpacs]{revtex4}

\usepackage{graphicx}
\usepackage{dcolumn}
\usepackage{bm}
\usepackage{graphics,pstricks,amsmath}
\usepackage{epsfig}

\begin{document}

\title{Prospects for sympathetic cooling of polar molecules: \\
NH with alkali-metal and alkaline-earth atoms -- a new hope}

\author{Pavel Sold\'{a}n\footnote{corresponding author}}
\email{pavel.soldan@mff.cuni.cz}

\affiliation{Department of Chemical Physics and Optics, Faculty
of Mathematics and Physics, Charles University in Prague, Ke
Karlovu 3, CZ-12116 Prague 2, Czech Republic}

\author{Piotr S. \.{Z}uchowski}
\email{piotr.zuchowski@durham.ac.uk}

\affiliation{Department of Chemistry, University of Durham,
South Road, Durham, DH1 3LE, United Kingdom}

\author{Jeremy M. Hutson}
\email{J.M.Hutson@durham.ac.uk}

\affiliation{Department of Chemistry, University of Durham,
South Road, Durham, DH1 3LE, United Kingdom}

\begin{abstract}
We explore the potential energy surfaces for NH molecules
interacting with alkali-metal and alkaline-earth atoms using
highly correlated {\em ab initio} electronic structure
calculations. The surfaces for interaction with alkali-metal
atoms have deep wells dominated by covalent forces. The
resulting strong anisotropies will produce strongly inelastic
collisions. The surfaces for interaction with alkaline-earth
atoms have shallower wells that are dominated by induction and
dispersion forces. For Be and Mg the anisotropy is small
compared to the rotational constant of NH, so that collisions
will be relatively weakly inelastic. Be and Mg are thus
promising coolants for sympathetic cooling of NH to the
ultracold regime.
\end{abstract}


\date{\today}


\maketitle

\section{Introduction}

In recent years there has been growing interest in the
production and properties of cold molecules. Possible
applications, such as controlled ultracold chemistry
\cite{Krems:PCCP:2008}, quantum information and computing
\cite{DeMille:2002}, and high-precision measurements of the
time-dependence of fundamental `constants' \cite{Hudson:2002,
vanVeldhoven:2004, Zelevinsky:2008}, make cold molecules
extremely interesting across many different fields of physics.

Two main approaches to the production of cold molecules can be
distinguished. One approach is based on the coherent formation
of ultracold molecules such as Cs$_{2}$ or RbCs in trapped
ultracold atomic gases \cite{Hutson:IRPC:2006}. The molecules
may be formed either by photoassociation \cite{Jones:RMP:2006}
or by Feshbach resonance tuning \cite{Kohler:RMP:2006}. They
inherit the $\mu$K-nK temperatures of the parent ultracold
atomic cloud and usually need very little further cooling.
Efforts in this area have led to the Bose-Einstein condensation
of Feshbach molecules \cite{Jochim:Li2BEC:2003, Zwierlein:2003,
Greiner:2003} and to the transfer of Feshbach molecules to
low-lying states \cite{Danzl:v73:2008, Ni:KRb:2008,
Lang:ground:2008}. There have also been considerable successes
in direct photoassociation to produce low-lying states
\cite{Sage:2005, Hudson:PRL:2008, Viteau:2008, Deiglmayr:2008}.

In the other approach, represented for example by Stark
deceleration \cite{Bethlem:IRPC:2003, Bethlem:2006} or helium
buffer-gas cooling \cite{Weinstein:CaH:1998}, preexisting
molecules are decelerated either by external fields or by
collisions with other particles and trapped in electrostatic or
magnetic traps. The temperature of the resulting molecular
cloud is usually in the K-mK region, and therefore new ways for
cooling the molecules further are being sought. A promising
route to cooling decelerated molecules down to the $\mu$K
region is offered by sympathetic cooling.

Sympathetic cooling, in which one species is cooled by thermal
contact with another much colder species, was originally
developed as a cooling technique for trapped ions
\cite{Larson:1986}. Diatomic \cite{Blythe:2005} and polyatomic
\cite{Ostendorf:2006} molecular ions have been cooled to
sub-Kelvin temperatures by thermal contact with cold ions, and
sympathetic cooling is expected to be capable of cooling ions
of very high mass, including those of biological relevance.
Sympathetic cooling has also been successful in producing
ultracold neutral atoms of species that are not themselves
suitable for evaporative cooling; for example it was used to
create the first Bose-Einstein condensates of potassium
$^{41}$K \cite{Modugno:2001}.

Sympathetic cooling is effective only if the rate of elastic
collisions is very large compared to the rate of inelastic
collisions. Elastic collisions exchange kinetic energy between
molecules and allow thermalization. However, inelastic
collisions in which internal energy is converted into relative
kinetic energy cause trap loss (if the energy released is
greater than the trap depth) or heating (if the energy released
is less than the trap depth). Magnetic and electrostatic traps
always trap molecules in low-field-seeking states, which are
not in their lowest state in the applied field. Since typical
traps have depths in the $\mu$K or mK range, most inelastic
collisions cause trap loss. A commonly stated rule of thumb is
that the ratio of elastic to inelastic collision rates must be
at least 100 for effective sympathetic cooling.

The obvious {\em coolants} for the sympathetic cooling of cold
molecules are alkali-metal atoms, which can be cooled to
ultra-low temperatures on demand. Sympathetic cooling of
photoassociated alkali-metal dimers in triplet states by
alkali-metal atoms was discussed by Cvita\v{s} \textit{et al.}\
\cite{Cvitas:hetero:2005, Cvitas:li3:2007}, who concluded that
the dimers would need to be in their ground rovibrational state
because of the unfavorable ratio of inelastic to elastic cross
sections in the sub-mK temperature region.

Sympathetic cooling of decelerated molecules by alkali-metal
atoms was first considered by Sold\'{a}n and Hutson
\cite{Soldan:2004}, who studied the interactions of rubidium
atoms with NH molecules. They showed that the $^{2}A''$ and
$^{4}A''$ states of RbNH (bound by covalent and dispersion
forces) are crossed by much deeper $^{2}A'$ and $^{2}A''$
ion-pair states in the energetically allowed region at linear
geometries. They concluded that the ion-pair states are likely
to have important consequences for the physics of sympathetic
cooling of molecules such as CH, NH and OH, because they
provide additional mechanisms for inelastic collisions and
three-body recombination.

Lara \textit{et al.}\ \cite{Lara:PRL:2006, Lara:PRA:2007}
subsequently focused on the interaction of OH molecules with
ultracold Rb atoms. They developed full sets of coupled
potential energy surfaces and carried out quantum collision
calculations including spin-orbit and hyperfine coupling. Once
again they found a deep ion-pair state ($^1A'$ for RbOH) that
crossed the covalent states at energetically accessible
geometries. However, even when the ion-pair state was excluded
from the calculation, the anisotropy of the potential for the
covalent states was enough to cause strong inelastic collisions
that would prevent sympathetic cooling except for atoms and
molecules in their absolute ground states. Lara \textit{et
al.}\ concluded in general that (i) light atomic partners are
desirable as coolants, because the resulting high centrifugal
barriers would suppress many inelastic channels; (ii) weak
coupling of the electron to the axis, which occurs in Hund's
case b molecules such as NH or CaH, would be beneficial; (iii)
the anisotropy of the atom-molecule surface should be
comparable to or smaller than the rotational constant of the
molecule; and (iv) closed-shell coolants, such as
alkaline-earth atoms, may produce more isotropic potential
energy surfaces than open-shell coolants, such as alkali-metal
atoms. Later a detailed \textit{ab initio} study of 2D
adiabatic potential energy surfaces for NH interacting with Rb
and Cs atoms was reported by Tacconi \textit{et al.}\
\cite{Tacconi:a:2007}, followed by studies of the quantum
dynamics of ultra-low-energy collision processes
\cite{Tacconi:b:2007,Tacconi:c:2007}.

Another set of potential coolants for sympathetic cooling are
alkaline-earth atoms. Calcium \cite{Binnewies:2001} and
strontium \cite{Mukaiyama:2003} atoms can be cooled and trapped
at temperatures of the order of $\mu$K. Mehlst\"{a}ubler {\em
et al.}\ \cite{Mehlstaubler:2008} have recently succeeded in
cooling magnesium atoms to sub-Doppler temperatures of 500
$\mu$K. To our knowledge no attempt has yet been made to cool
Be atoms.

Very recently, \.{Z}uchowski and Hutson \cite{Zuchowski:NH3:2008}
surveyed interactions of NH$_{3}$ molecules with alkali-metal and
alkaline-earth atoms. All the systems exhibited deep potential wells
(890 to 5100 cm$^{-1}$) when the atom was on the N side of the
molecule, and shallow potential wells (100 to 130 cm$^{-1}$) when
the atom was on the H side, resulting in very strong anisotropy of
the surfaces. This will produce strong inelasticity in the molecular
rotational and inversion degrees of freedom and sympathetic cooling
is unlikely to be successful for molecules in low-field-seeking
states.


In this paper we survey the possibilities for sympathetic
cooling of NH molecules, which have very recently been cooled
and magnetically trapped at 0.7 K in their ground
$X^{3}\Sigma^{-}$ state \cite{Campbell:2007} by buffer-gas
cooling. NH molecules in their metastable $a^{1}\Delta$ state
have also been Stark-decelerated \cite{Hoekstra:2007} and
electrostatically trapped at temperatures of 60-100 mK, and
there is a proposal \cite{vandeMeerakker:2001} to transfer
$a^{1}\Delta$ molecules to the $X^{3}\Sigma^{-}$ state. In the
present paper we investigate the interactions of
NH($X^{3}\Sigma^{-}$) molecules not only with all the relevant
alkali-metal (Alk) atoms, but also for the first time with the
alkaline-earth (Ae) atoms. We characterize the potential energy
surfaces of the covalent and dispersion-bound states of the
AlkNH and AeNH systems and locate their conical intersections
with the ion-pair states. We show that Be and Mg atoms are
promising candidates for sympathetic cooling of
NH($X^{3}\Sigma^{-}$) molecules.

\section{Methods}

To facilitate future quantum dynamics calculations, all results
are reported in Jacobi coordinates. The intermolecular distance
$R$ is the distance between the alkali-metal or alkaline-earth
atom and the center of mass of the NH molecule. The angle
$\theta$ is measured at the center of mass and is zero for
linear atom-HN geometries. In all our calculations the NH bond
length $r$ is fixed at the experimentally determined
equilibrium value for the free monomer, 1.0367 \AA\
\cite{Brazier:1986}.

Supermolecular coupled-cluster calculations were carried out
using a single-reference restricted open-shell variant
\cite{Knowles:1993} of the coupled cluster method
\cite{Cizek:1966} with single, double and non-iterative triple
excitations [RCCSD(T)]. All electrons from the ``outer-core''
orbitals (1$s$, 2$s$2$p$, 3$s$3$p$, 4$s$4$p$, and 5$s$5$p$ for
Li and Be, Na and Mg, K and Ca, Rb and Sr, and Cs,
respectively) were included in the RCSSD(T) calculations. All
the \textit{ab initio} calculations were performed using the
MOLPRO package \cite{molpro_brief:2006}.

To describe the interaction between the outer-core and valence
electrons, and to reduce basis-set superposition errors, rather
large basis sets are needed. We use the correlation-consistent
polarized valence quintuple-$\zeta$ (cc-pV5Z) basis sets of
Dunning \cite{Dunning:1989} for hydrogen (without the $g$
functions) and for nitrogen (without the $h$ functions). Both
these basis sets were augmented in an even-tempered manner and
used in uncontracted form. For lithium, beryllium, sodium,
magnesium, potassium, and calcium atoms, we use the
correlation-consistent polarized core-valence quintuple-$\zeta$
cc-pCV5Z basis sets of Iron \textit{et al.}\ \cite{Iron:2003},
again without the $h$ functions. The Li, Be, Na, and Mg basis
sets were used in fully uncontracted form and those for K and
Ca were used partially contracted. These basis sets were also
augmented by additional even-tempered diffuse functions.  The
resulting aug-cc-pCV5Z basis sets consist of
(15$s$,10$p$,8$d$,6$f$,4$g$), (20$s$,13$p$,9$d$,7$f$,5$g$), and
[13$s$,12$p$,9$d$,7$f$,5$g$] functions for Li, Na, and K,
respectively, and of (15$s$,9$p$,8$d$,6$f$,4$g$),
(21$s$,15$p$,9$d$,7$f$,5$g$), and [13$s$,12$p$,9$d$,7$f$,5$g$]
functions for Be, Mg, and Ca, respectively. For rubidium,
strontium, and cesium, we use the small-core scalar
relativistic effective core potentials ECP28MDF and ECP46MDF
\cite{Lim:2005, Lim:2006}, together with the corresponding
valence basis sets. These basis sets were augmented in the
even-tempered manner and used in uncontracted form. The
resulting basis sets consisted of (14$s$,11$p$,6$d$,4$f$,2$g$),
(15$s$,12$p$,7$d$,5$f$,2$g$), and (13$s$,12$p$,6$d$,4$f$,3$g$)
primitive Gaussian functions for Rb, Sr, and Cs, respectively.

All interaction energies are calculated with respect to the
separated-atom-molecule limit, with both the atom and the
molecule in their ground states. The full counterpoise
correction of Boys and Bernardi \cite{Boys:1970} is used to
compensate for basis set superposition errors. Optimizations of
the counterpoise-corrected dimer interaction energies are
performed making use of a general optimization algorithm
implemented in MOLPRO.

In order to describe the dispersion-bound state of MgNH for all
geometries, we use a version of symmetry-adapted perturbation
theory (SAPT) based on a density-functional theory (DFT)
description of the monomers. In the SAPT(DFT) method
\cite{Zuchowski:SAPT-DFT:2008} (which we use here only for
MgNH), the interaction energy is obtained as a sum of
contributions,
\begin{eqnarray}
E^{\rm SAPT(DFT)}_{\rm int} &=& E^{(1)}_{\rm elst} +
E^{(1)}_{\rm exch}+ E^{(2)}_{\rm disp}  + E^{(2)}_{\rm ind}  \nonumber \\
&+& E^{(2)}_{\rm exch-disp} + E^{(2)}_{\rm exch-ind},
\end{eqnarray}
where $E^{(1)}_{\rm elst}$ is the electrostatic energy,
$E^{(1)}_{\rm exch} $ is the first-order exchange energy,
$E^{(2)}_{\rm disp} $ and $ E^{(2)}_{\rm ind} $ are the
second-order dispersion and induction energies, and
$E^{(2)}_{\rm exch-disp}$ and $E^{(2)}_{\rm exch-ind} $ are
their exchange counterparts. The first-order terms are
calculated using Kohn-Sham orbitals, while the dispersion,
induction and exchange-induction terms are evaluated using
coupled Kohn-Sham density susceptibilities. The
exchange-dispersion term is estimated as described in Ref.\
\cite{Zuchowski:SAPT-DFT:2008}. However, in Ref.\
\cite{Zuchowski:SAPT-DFT:2008} the second-order exchange
corrections are given in the so-called $S^2$ approximation,
which neglects terms of third and higher powers in the overlap
matrix $S$. Since the overlap between Mg and NH is large, we
scale the second-order exchange corrections by $E^{(1)}_{\rm
exch}(S^2) / E^{(1)}_{\rm exch}$. This procedure was introduced
by Patkowski {\em et al.}\ \cite{Patkowski:SAPT-DFT:2007} to
improve the performance of SAPT for systems with very diffuse
monomer densities.

Our SAPT calculations use the PBE0 density functional
\cite{Adamo:1999}, with augmented correlation-consistent
polarized valence quadruple-zeta (aug-cc-pVQZ) basis sets
\cite{Iron:2003, Dunning:1989} supplemented with bond functions
0.9,0.3,0.1 $s$ and $p$, 0.6,0.2 $d$ and $f$ placed at the
midpoint between the Mg atom and the center of mass of NH. The
Tozer-Handy asymptotic correction \cite{Tozer:1998} of the
exchange-correlation potential is used, with splicing
parameters 3.5 and 4.7~\AA.

\section{Results and discussion}

\subsection{Alkali-metal atom + NH interactions}

In discussing the electronic structure of NH interacting with
alkali metals (Alk), it is convenient to begin with linear
arrangements. At linear geometries (point group $C_{\infty
v}$), there are two covalent states $^{4}\Sigma^{-}$ and
$^{2}\Sigma^{-}$, which correlate with the Alk($^{2}S$) +
NH($^{3}\Sigma^{-}$) dissociation limit. These are crossed by
an ion-pair $^{2}\Pi$ state, which in a diabatic representation
correlates with the Alk$^{+}$($^{1}S$) + NH$^{-}$($^{2}\Pi$)
dissociation limit. (In an adiabatic representation this state
changes configuration at a long-range avoided crossing with a
higher $^{2}\Pi$ state, and in the new configuration it
correlates with Alk($^{2}P$) + NH($^{3}\Sigma^{-}$); the
ion-pair configuration is then carried up by a cascade of
similar avoided crossings until it reaches the
Alk$^{+}$($^{1}S$) + NH$^{-}$($^{2}\Pi$) dissociation limit).
At non-linear geometries (point group $C_{s}$), the
$^{4}\Sigma^{-}$ and $^{2}\Sigma^{-}$ states become $^{4}A''$
and $^{2}A''$ states, while the $^{2}\Pi$ state is subject to
the Renner-Teller effect and splits into two states with the
electron hole either in the triatomic plane ($^{2}A'$) or
perpendicular to it ($^{2}A''$). In cuts through the potential
at fixed N-H distance, the covalent $^{2}A''$ and ion-pair
$^{2}A''$ states avoided-cross at nonlinear geometries but form
a conical intersection at linear geometries. In the full
three-dimensional picture they form a seam of conical
intersections parameterized by the N-H distance.

\begin{figure}[tbph]
\begin{center}
\includegraphics[width=0.65\linewidth,angle=270]{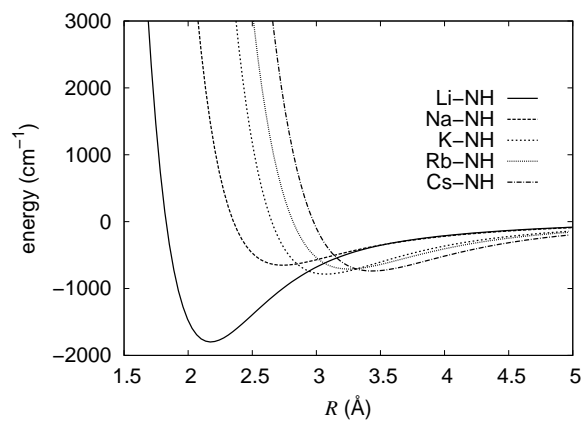}
\caption{One-dimensional cuts through the lowest $^{4}\Sigma^-$
potential energy surfaces of AlkNH  systems  at the linear
Alk-NH arrangement. \label{alknhfig}}
\end{center}
\end{figure}

We consider first the quartet states. The potential curves for
the $^{4}\Sigma^{-}$ states at linear Alk-NH geometries are
shown in Fig.\ \ref{alknhfig}. Results from geometry
optimization are given in Table \ref{alknh} for both linear
configurations.

\begin{table*}[tbpb]
\caption{Lowest $^{4}\Sigma^{-}$ and $^{2}\Pi$ states of linear
AlkNH: minima ($R_{\rm min}$, $V_{\rm min}$) and crossing
points ($R_{\rm x}$, $V_{\rm x}$) at different arrangements
(Alk-NH, Alk-HN). Energies are given in cm$^{-1}$ and distances
in \AA.} \label{alknh}
\begin{ruledtabular}
\begin{tabular}{lrrrrrrrrr}
 & \multicolumn{6}{c}{Alk-NH} & & \multicolumn{2}{c}{Alk-HN}  \\
\cline{2-7} \cline{9-10}
Alk & $R_{\rm min}^{\Sigma}$ & $V_{\rm min}^{\Sigma}$ & $R_{\rm x}$ & $V_{\rm x}$ & $R_{\rm min}^{\Pi}$ & $V_{\rm min}^{\Pi}$ & & $R_{\rm min}^{\Sigma}$ & $V_{\rm min}^{\Sigma}$ \\
\colrule
Li & 2.176 & -1799.1 & 3.09 & -600 & 1.78 & -21073 & & 4.658 & -115.3 \\
Na & 2.737 &  -651.3 & 3.21 & -470 & 2.13 & -13287 & & 4.872 &  -98.9 \\
K  & 3.073 &  -784.7 & 3.84 & -432 & 2.42 & -14235 & & 5.386 &  -91.1 \\
Rb & 3.254 &  -709.3 & 3.99 & -412 & 2.53 & -13918 & & 5.521 &  -87.3 \\
Cs & 3.435 &  -737.9 & 4.31 & -381 & 2.65 & -15116 & & 5.761 &  -85.0 \\
\end{tabular}
\end{ruledtabular}
\end{table*}

From the viewpoint of sympathetic cooling, the most important
feature of the potential surface is its anisotropy. For sympathetic
cooling to be successful for  low-field-seeking molecular states, we
need a potential surface where the anisotropy is small compared to
the rotational constant of the monomer ($B_e = 16$ cm$^{-1}$ for
NH). The global minimum for quartet states is at linear Alk-NH
geometries for all systems. For LiNH, the well depth at the global
minimum is about 1800 cm$^{-1}$, while at the secondary minimum
(Li-HN) it is only 115 cm$^{-1}$. The absolute well depths for the
other AlkNH systems are a factor of 2 to 3 smaller than for Li-NH,
while the depths of the secondary wells are comparable for all the
alkali-metal atoms. Nevertheless, for all the AlkNH systems the
anisotropy is very large compared to the rotational constant of NH.

The large anisotropy of the AlkNH quartet surfaces results from
strong $sp$ mixing of the alkali-metal orbitals in the Alk-NH
arrangement. Such mixing is much weaker in the Alk-HN
arrangement. To quantify this we have performed Mulliken
population analysis \cite{Mulliken:1955}. The strongest effect
is observed for LiNH, where the partial occupancy of the
valence $p_{z}$ orbital is 0.086 in the Li-NH arrangement
compared to 0.004 in the Li-HN arrangement. The $p_z$ occupancy
for Alk-NH arrangements decreases down the periodic table: Na
0.068, K 0.037, Rb 0.027 and Cs 0.021. This hybridization
tendency is similar to that found for the alkali-metal trimers
in the quartet states \cite{Soldan:2003}.

The present calculations give a quartet well depth for RbNH
that is about 12\% deeper than the value of 78 meV (630
cm$^{-1}$) obtained by Sold\'{a}n and Hutson \cite{Soldan:2004}
using multireference configuration interaction (MRCI)
calculations. The present work used RCCSD(T) calculations,
which give a better treatment of dispersion effects, and also
used much larger basis sets. Our well depths are also
substantially deeper than those reported for Rb-NH and Cs-NH by
Tacconi \textit{et al.}\ \cite{Tacconi:a:2007} using methods
and basis sets similar to those of Ref.\ \cite{Soldan:2004}.

For all the Alk-NH systems there are also covalent states of
$^2\Sigma^-$ symmetry ($^2A''$ at bent geometries) and an
ion-pair state of $^2\Pi$ symmetry ($^2A'$ and $^2A''$ at bent
geometries). As described above, there is an avoided crossing
between the two $^{2}A''$ states. Table \ref{alknh} gives the
equilibrium positions and well depths of the $^2\Pi$ states for
Alk-NH geometries, and it may be seen that for all systems the
ion-pair well is more than 13000 cm$^{-1}$ deep. The potential
curve for the $^2\Sigma^-$ state cannot be obtained from
RCCSD(T) calculations, but it is qualitatively similar to that
for the $^4\Sigma^-$ state in the long-range region
\cite{Soldan:2004}. Table \ref{alknh} includes the position and
energy of the crossing point between the $^4\Sigma^-$ and
$^2\Pi$ curves, and it may be seen that the crossing is always
outside the minimum of the $\Sigma$ state. Because of this, the
lowest adiabatic surface of either $^{2}A'$ or $^{2}A''$
symmetry always has a very deep well of ion-pair character.
This well is strongly anisotropic, so that any collision that
samples the doublet surfaces is likely to be strongly
inelastic.

In conclusion, it appears that both the doublet and quartet states
of Alk-NH systems have sufficient anisotropy to prevent sympathetic
cooling for low-field-seeking molecular states.


\subsection{Alkaline-earth atom + NH interactions}

The potential energy surfaces for alkaline-earth atoms (Ae)
interacting with NH are substantially different from those for
AlkNH systems. At linear geometries, there is one
dispersion-bound state, $^{3}\Sigma^{-}$, which correlates with
the Ae($^{1}S$) + NH($^{3}\Sigma^{-}$) dissociation limit. This
state is crossed by ion-pair $^{3}\Pi$ and $^{1}\Pi$ states,
which in a diabatic representation correlate with the
Ae$^{+}$($^{2}S$) + NH$^{-}$($^{2}\Pi$) dissociation limit. At
non-linear geometries (point group $C_{s}$), the
$^{3}\Sigma^{-}$ state becomes a $^{3}A''$ state, and the
$^{3}\Pi$ state is subject to the Renner-Teller effect and
splits into two states with the electron hole either in the
triatomic plane ($^{3}A'$) or perpendicular to it ($^{3}A''$).
In cuts at fixed NH bond length, the dispersion-bound $^{3}A''$
and ion-pair $^{3}A''$ states form a conical intersection at
linear geometries, while in the full three-dimensional picture
they form a seam of conical intersections parameterized by the
N-H distance.

\begin{figure*}[tbph]
\includegraphics[width=0.47\linewidth]{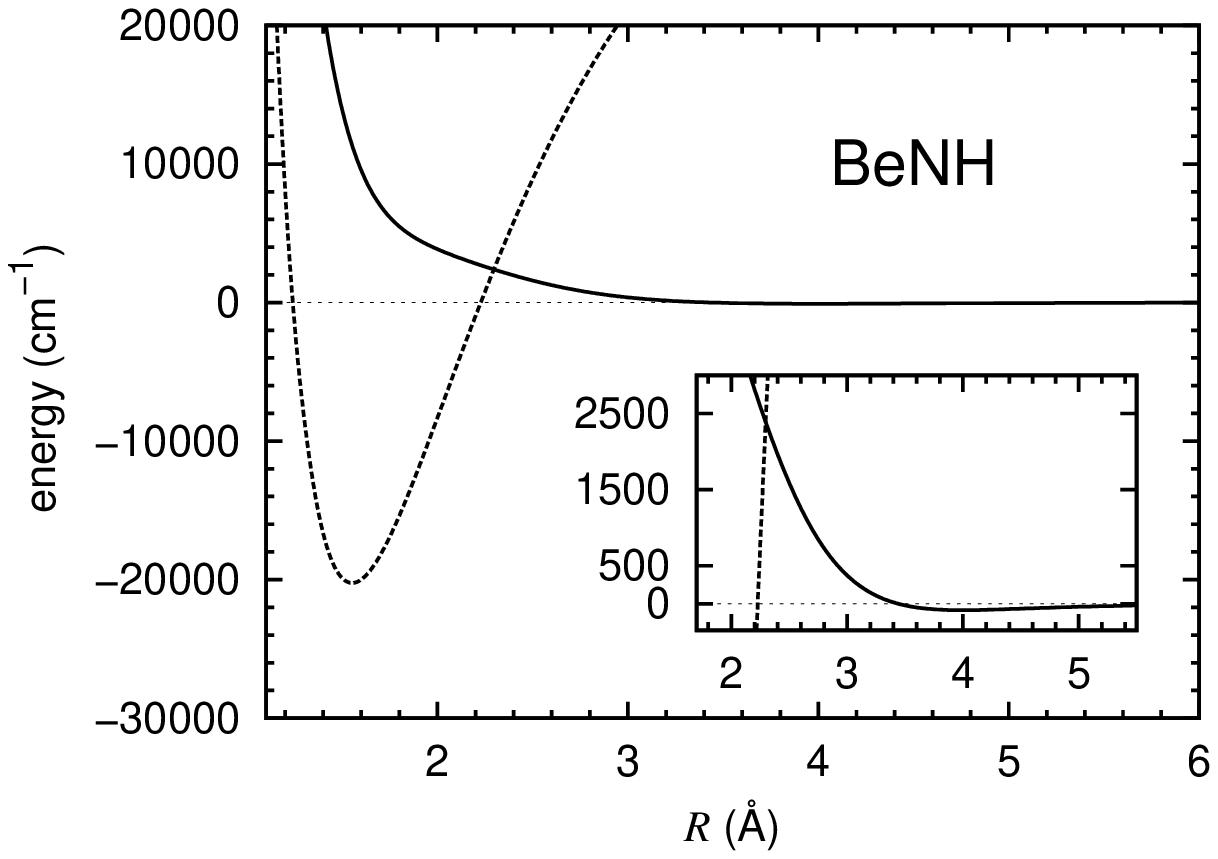}
\includegraphics[width=0.47\linewidth]{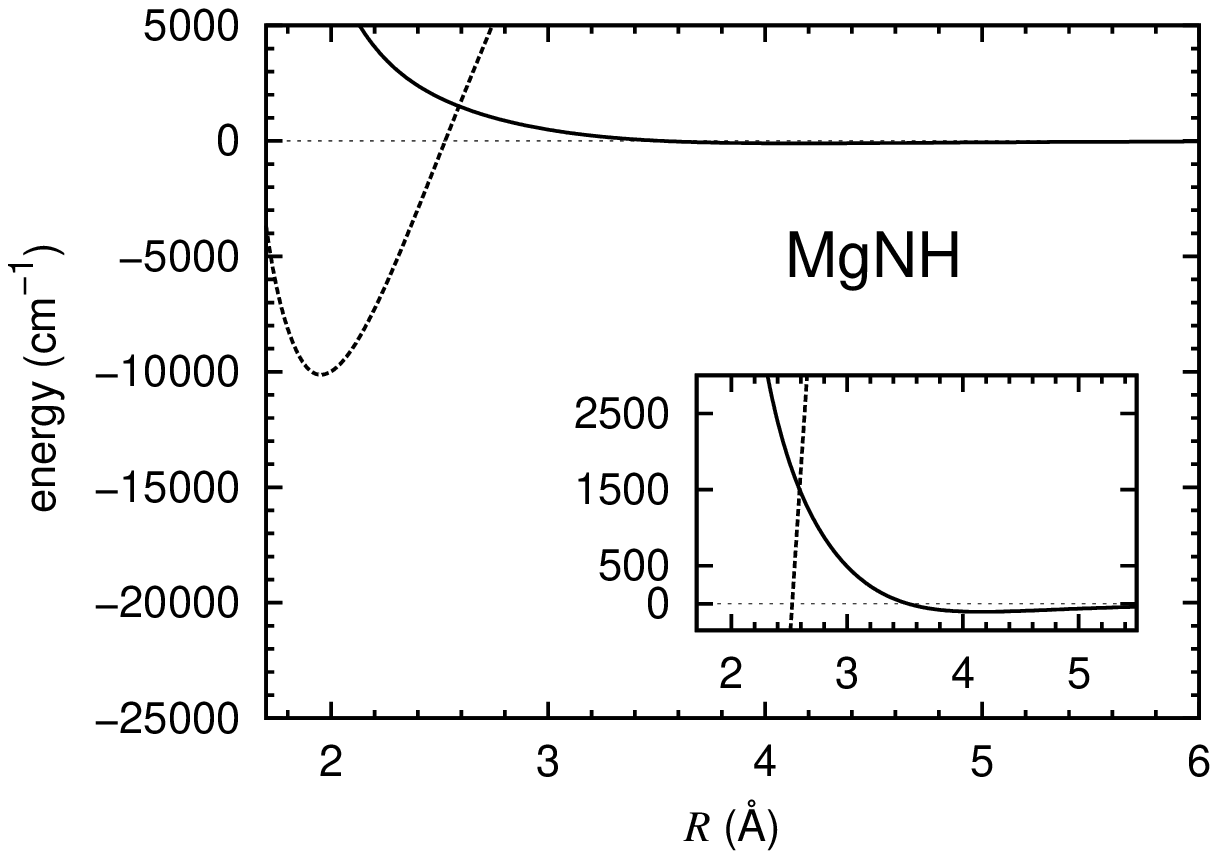}
\includegraphics[width=0.47\linewidth]{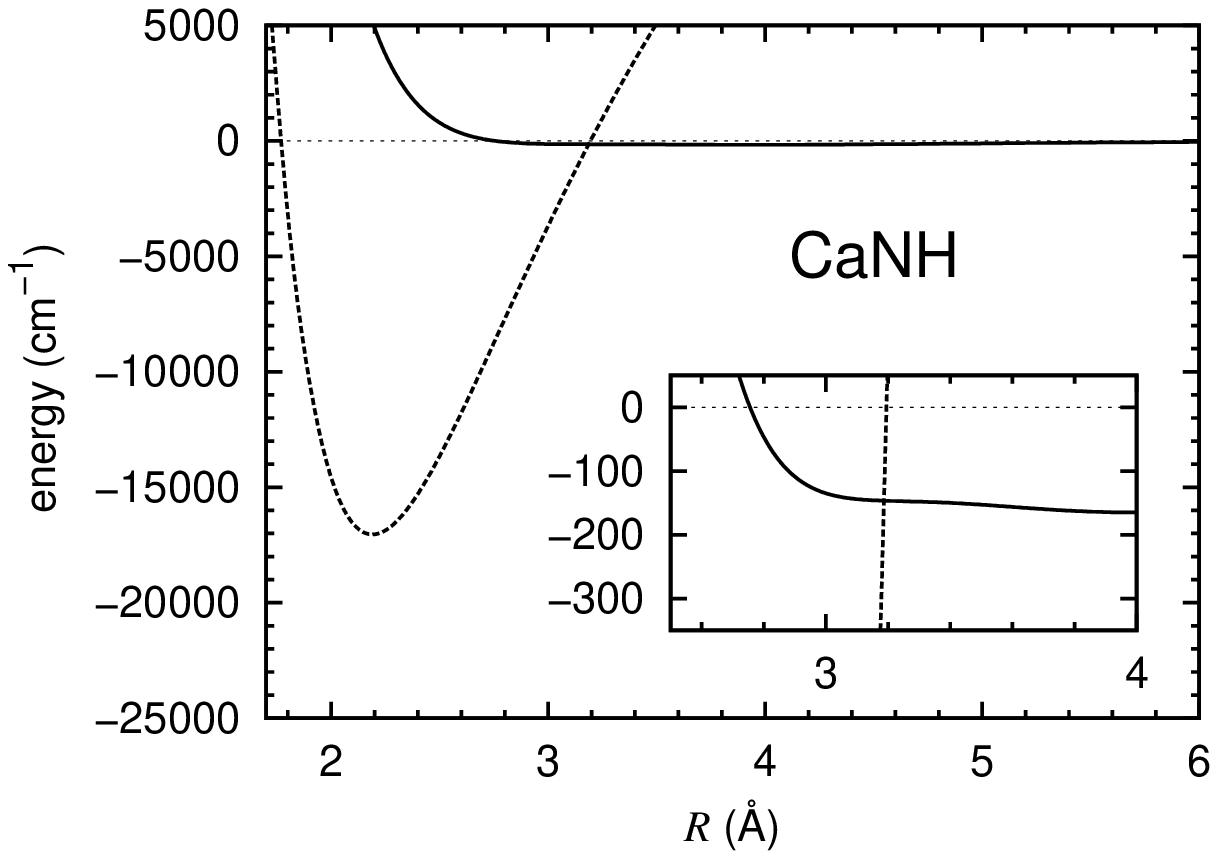}
\includegraphics[width=0.47\linewidth]{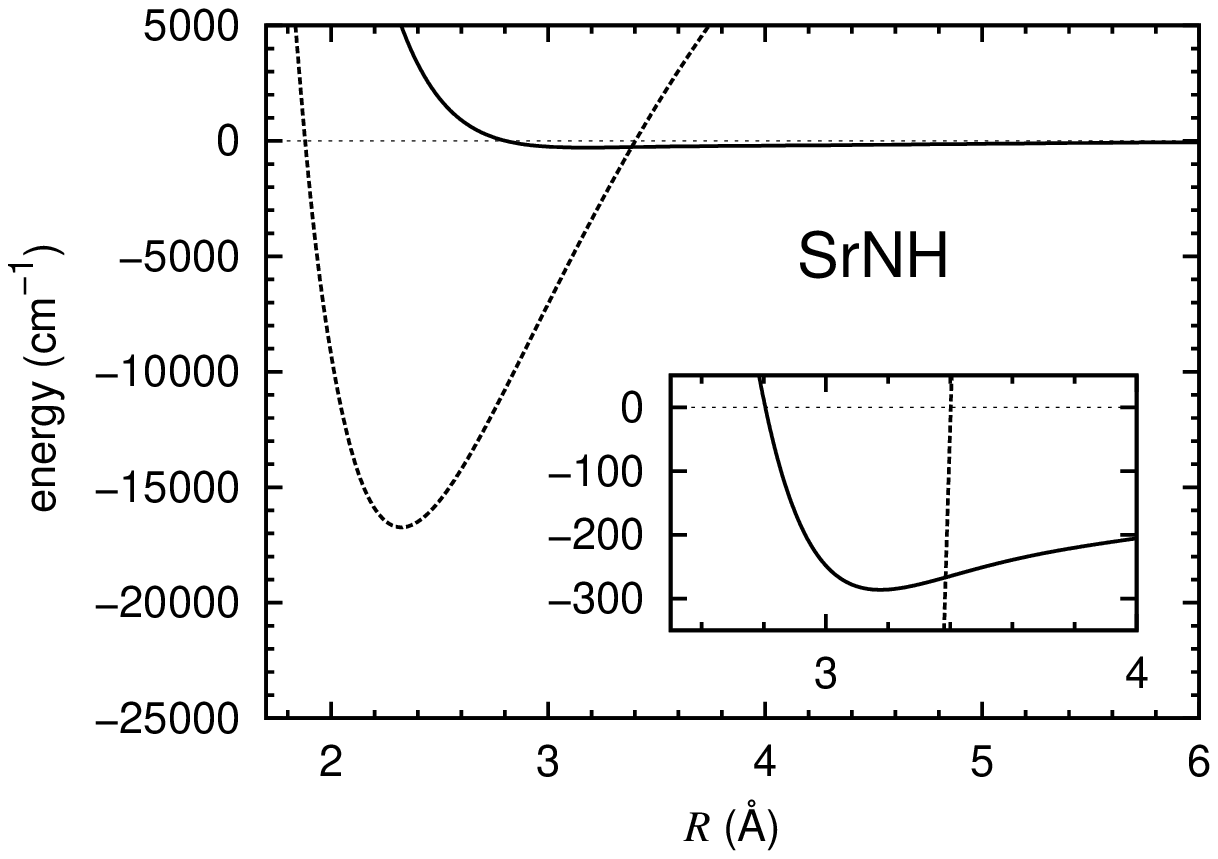}
\caption{One-dimensional cuts through the lowest $^{3}\Sigma^-$
(full) and $^{3}\Pi$ (dashed) potential energy surfaces of AeNH
systems at the linear Ae-NH arrangement. For clarity, the
region near the crossing of the dispersion-bound and ion-pair
states is magnified. } \label{aenhfig}
\end{figure*}

\begin{figure*}[tbph]
\includegraphics[width=0.47\linewidth]{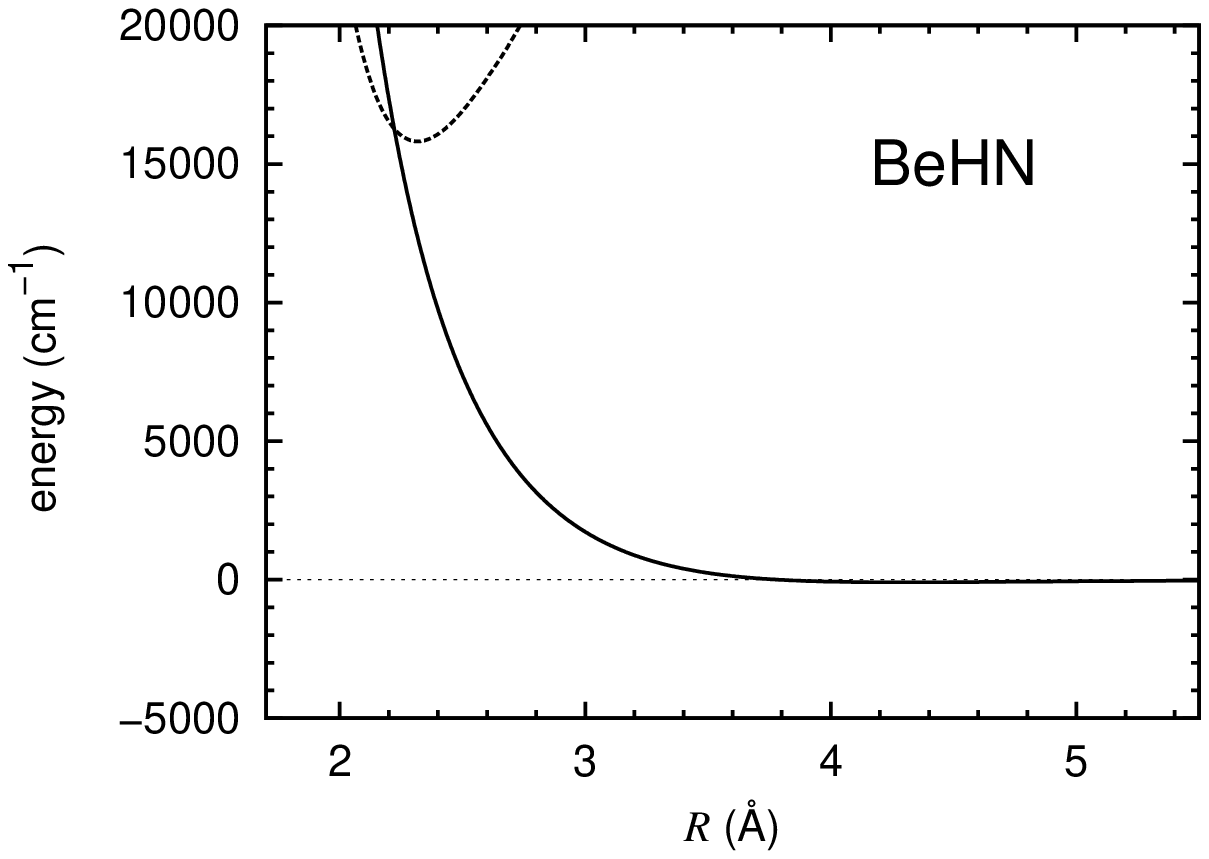}
\includegraphics[width=0.47\linewidth]{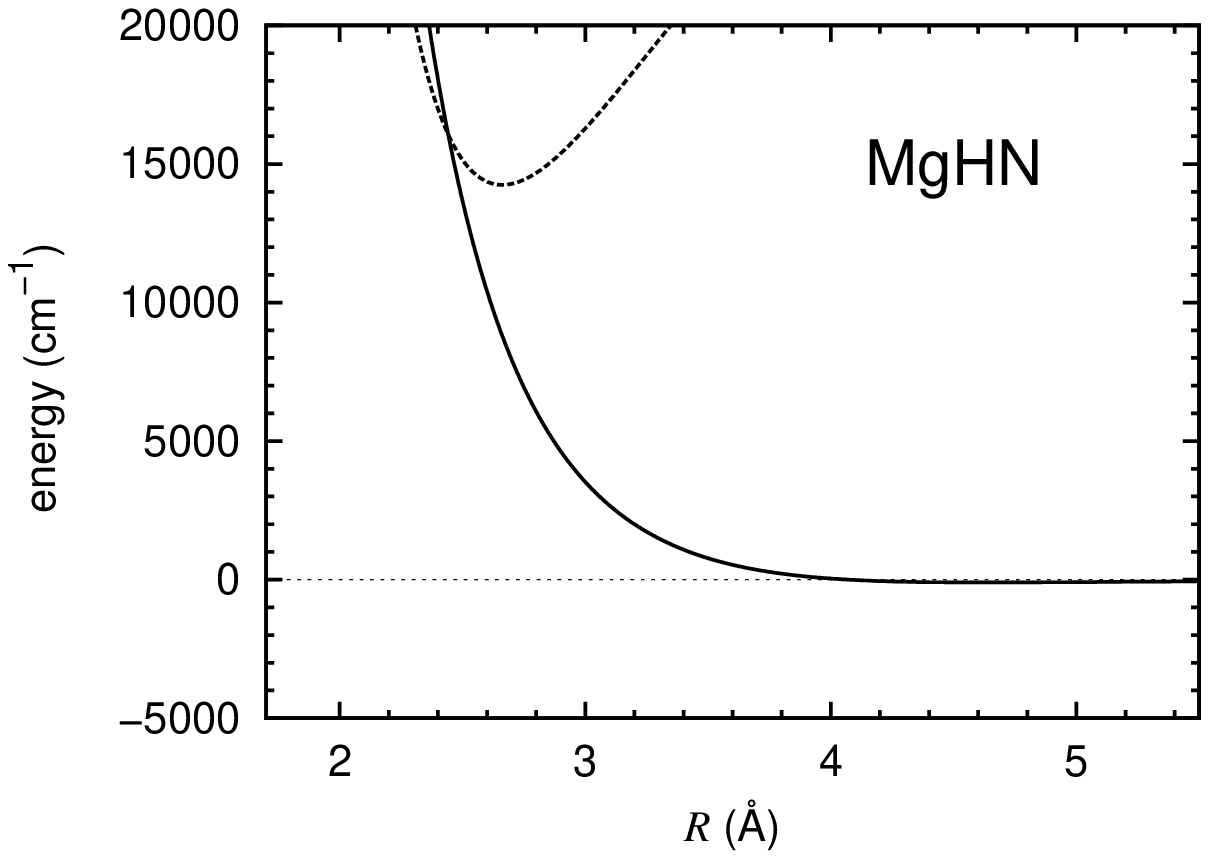}
\includegraphics[width=0.47\linewidth]{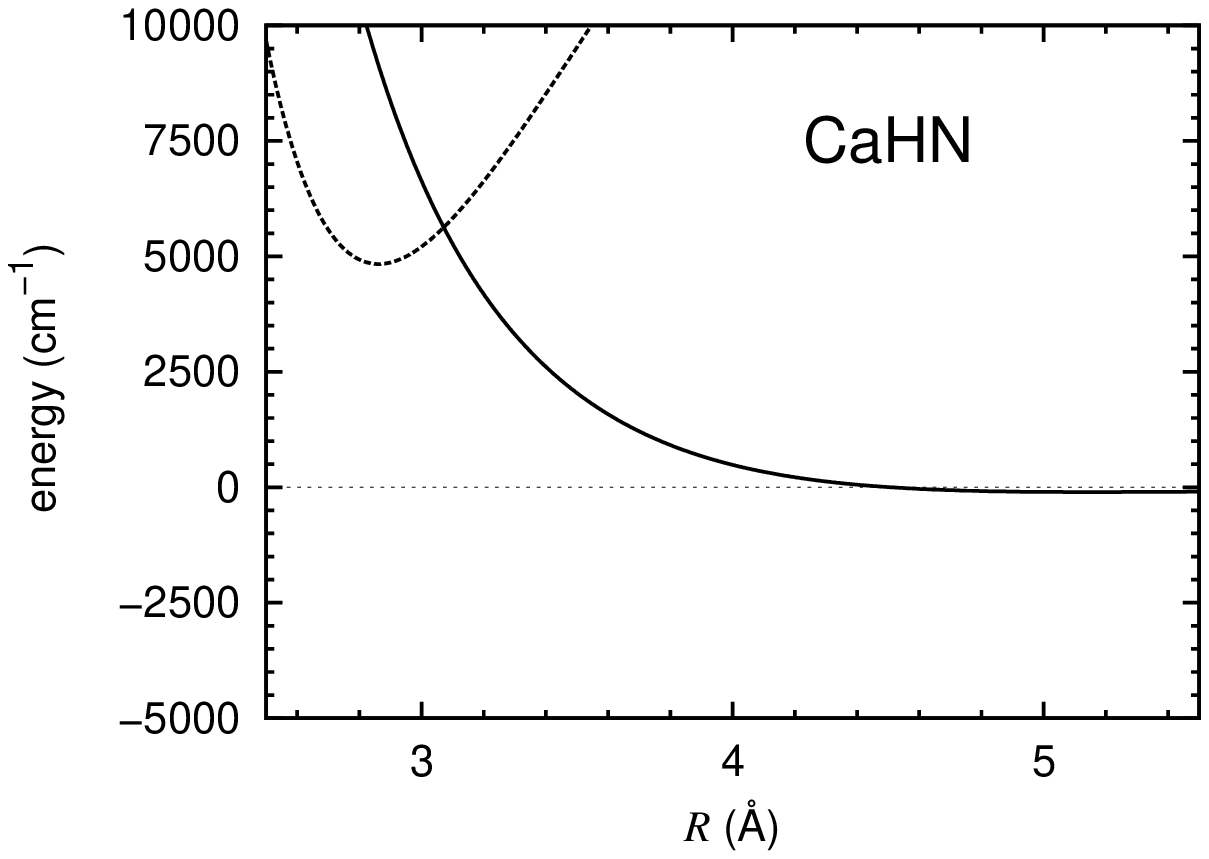}
\includegraphics[width=0.47\linewidth]{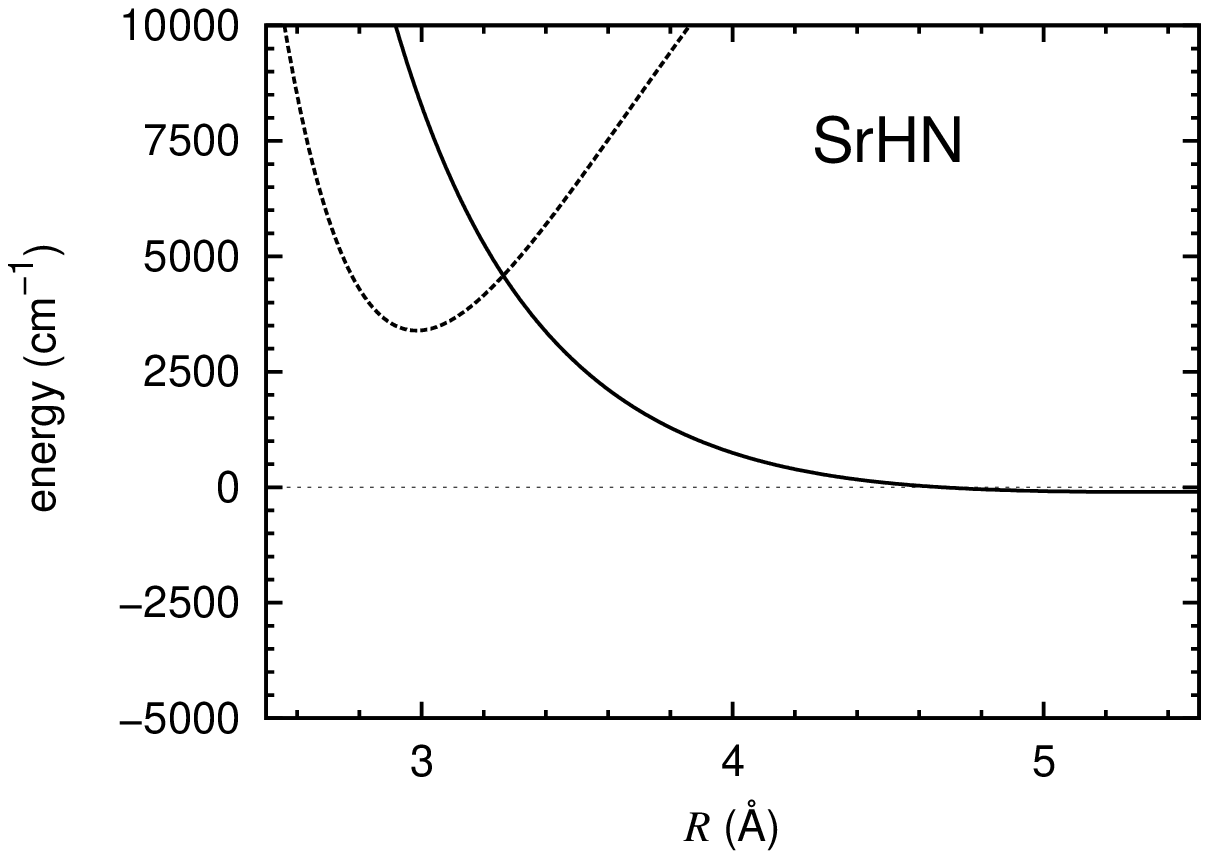}
\caption{One-dimensional cuts through the lowest $^{3}\Sigma^-$
(full) and $^{3}\Pi$ (dashed) potential energy surfaces of AeNH
systems  at the linear Ae-HN arrangement.} \label{aehnfig}
\end{figure*}

The counterpoise-corrected equilibrium distances and well
depths for AeNH systems are shown in Table \ref{aenh} for both
Ae-NH and Ae-HN linear geometries. The corresponding potential
curves are shown in Figs.\ \ref{aenhfig} and \ref{aehnfig} for
Ae-NH and Ae-HN geometries, respectively. It may be seen that
the anisotropy of the dispersion-bound state is considerably
smaller for the alkaline-earth atoms than for the alkali-metal
atoms. However, for Ca and Sr the difference between the well
depths at the two linear geometries (60.4 and 184.6 cm$^{-1}$,
respectively) is still several times the NH rotational constant
($B_e = 16$ cm$^{-1}$). For  Be and Mg, by contrast, the
difference is only 10.5 cm$^{-1}$ and 3.6 cm$^{-1}$,
respectively, which is {\em smaller} than the NH rotational
constant. The difference may be understood in terms of the
smaller $s$-$p$ excitation energies for Ca and Sr (1.9 and 1.8
eV, respectively) compared to those of Be and Mg (both 2.7 eV).

\begin{table*}[tr!]
\caption{Lowest $^{3}\Sigma^{-}$  and $^{3}\Pi$ states of
linear AeNH: minima ($R_{\rm min}$, $V_{\rm min}$) and crossing
points ($R_{\rm x}$, $V_{\rm x}$) at different arrangements
(Ae-NH, Ae-HN). Energies are given in cm$^{-1}$ and distances
in \AA.} \label{aenh}
\begin{ruledtabular}
\begin{tabular}{lrrrrrrrrr}
 & \multicolumn{6}{c}{Ae-NH} & & \multicolumn{2}{c}{Ae-HN}  \\
\cline{2-7} \cline{9-10}
Ae & $R_{\rm min}^{\Sigma}$ & $V_{\rm min}^{\Sigma}$ & $R_{\rm x}$ & $V_{\rm x}$ & $R_{\rm min}^{\Pi}$ & $V_{\rm min}^{\Pi}$ & & $R_{\rm min}^{\Sigma}$ & $V_{\rm min}^{\Sigma}$ \\
\colrule
Be & 3.995 &  -84.5 & 2.30 & 2390 & 1.55 & -20240 & & 4.301 &  -95.4 \\
Mg & 4.157 & -106.5\footnotemark[1] & 2.59 & 1510 & 1.95 & -10120 & & 4.636 & -103.0\footnotemark[2]\\
Ca & 3.963 & -165.7 & 3.19 & -146 & 2.19 & -17041 & & 5.149 & -104.5 \\
Sr & 3.175 & -286.4 & 3.39 & -267 & 2.32 & -16734 & & 5.340 & -101.8 \\
\end{tabular}
\end{ruledtabular}
\footnotetext[1]{$E^{\rm SAPT(DFT)}_{\rm int} = -113.4 $cm$^{-1}$
and corresponding $R_{\rm min}^{\Sigma}=4.24$ \AA   }
\footnotetext[2]{$E^{\rm SAPT(DFT)}_{\rm int} = -91.7$ cm$^{-1} $
and corresponding $R_{\rm min}^{\Sigma}=4.74$ \AA    }
\end{table*}

Another important feature is the position of the crossing
between the dispersion-bound $^{3}\Sigma^-$ state and the
ion-pair $^{3}\Pi$ states. These are also tabulated in Table
\ref{aenh} and shown in Figs.\ \ref{aenhfig} and \ref{aehnfig}.
For Be-NH and Mg-NH the crossing occurs fairly high on the
repulsive wall of the $^{3}\Sigma^{-}$ state, while for Ca-NH
and Sr-NH it occurs at negative energies (in the potential
well). This may be crucial for collisional properties. If the
crossing is located at negative energies, as for Ca-NH and
Sr-NH, the deep, strongly anisotropic ion-pair well may be
accessed in low-energy collisions and is likely to result in
strong inelasticity. On the other hand, if the crossing occurs
at a high energy in a classically inaccessible region, as for
Be-NH and Mg-NH, the deep ion-pair well is accessible only by
tunneling through a barrier and may not have a strong effect on
collisions.

For Ae-HN geometries, shown in Table \ref{aenh}, the crossing occurs
high on the repulsive wall for all the AeNH systems.

\subsection{Mg-NH interaction potential }

As shown above, the BeNH and MgNH systems have potential energy
surfaces that appear promising for sympathetic cooling.
However, Mg has been successfully laser-cooled
\cite{Mehlstaubler:2008} whereas Be has not. We therefore focus
in this section on developing a complete potential energy
surface for interaction of Mg with NH($^3\Sigma^-$).

For bent geometries near the conical intersection, the two
lowest triplet states of $A''$ symmetry are near-degenerate.
Under these circumstances single-reference coupled-cluster
calculations are inappropriate. One alternative, which we have
previously applied for RbOH \cite{Lara:PRL:2006,
Lara:PRA:2007}, is to carry out multireference configuration
interaction calculations including single and double
excitations (MR-CISD). However, for MgNH the contribution from
triple excitations is extremely large: for example, the well
depths of the linear $^3\Sigma^-$ state is underestimated by
40\% in RCCSD calculations. Because of this, we use SAPT(DFT)
calculations to study nonlinear configurations of MgNH. It has
recently been demonstrated \cite{Patkowski:SAPT-DFT:2007,
Zuchowski:SAPT-DFT:2008} that SAPT(DFT) gives reasonably good
results for Mg$_2$, NH-He and MgHe, and the polarizabilities
and Van der Waals coefficients for Mg$_2$ are reproduced with
an accuracy of a few percent. A further advantage of the
perturbation theory is that, by starting from zeroth-order
wavefunctions corresponding to neutral monomers, we produce
diabatic potential energy surfaces corresponding to neutral
Mg-NH without contamination from ion-pair states.

\begin{figure}[tbph]
\includegraphics[width=0.9\linewidth]{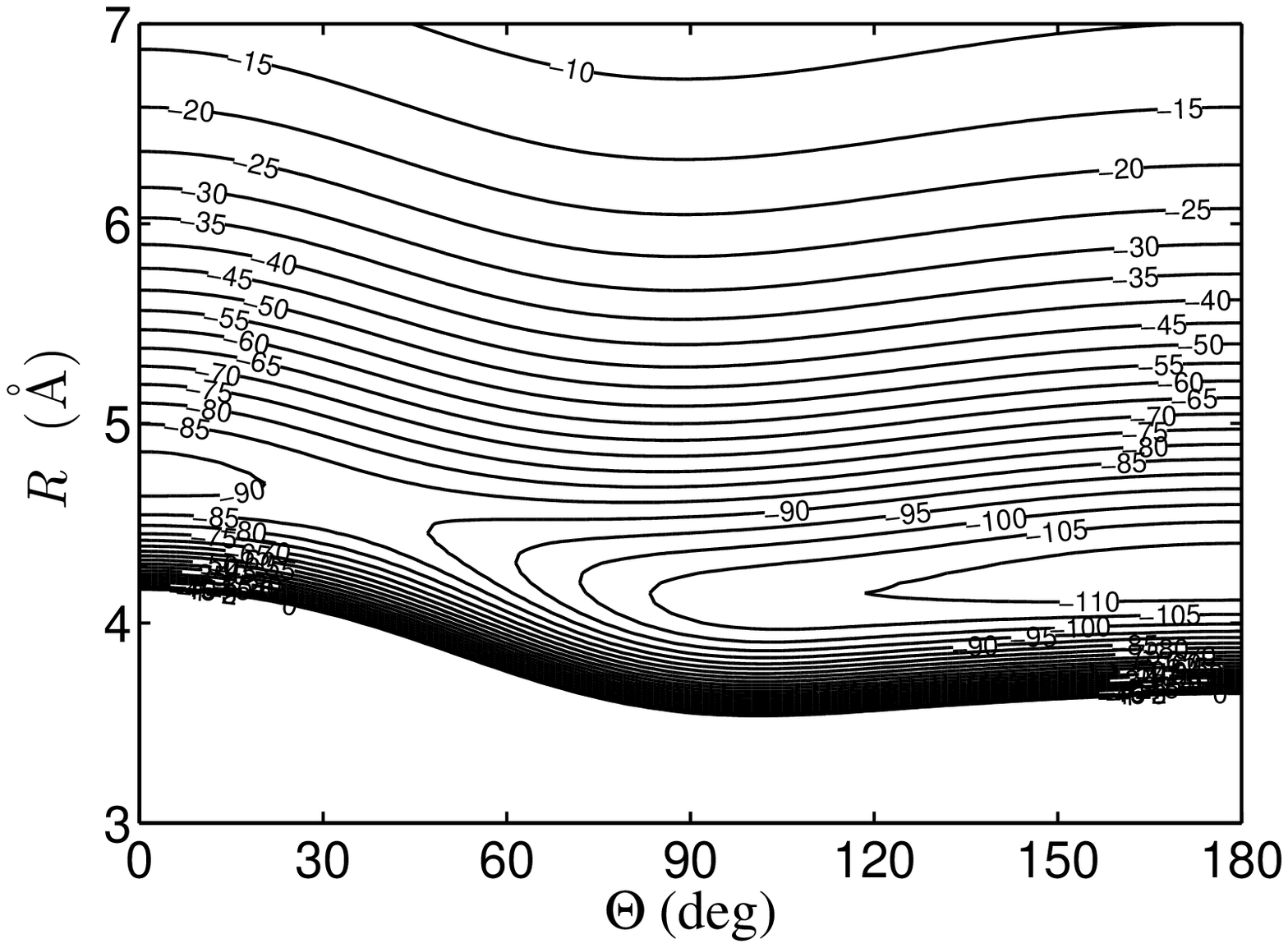}
\caption{Contour plot of the potential energy surface for the
dispersion-bound state of Mg-NH, from SAPT(DFT) calculations (in
cm$^{-1}$).} \label{figmgnh}
\end{figure}

\begin{figure}[tbph]
\includegraphics[width=0.9\linewidth]{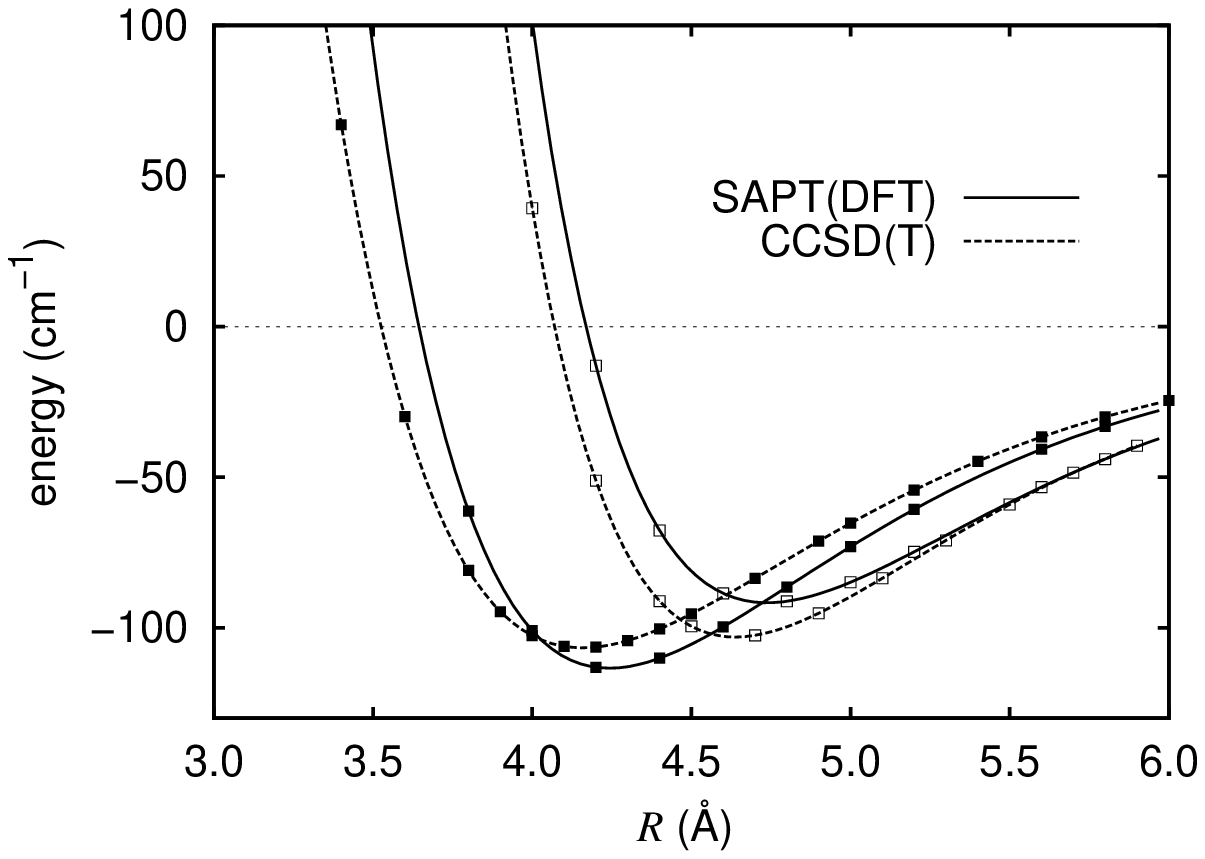}
\caption{Comparison of SAPT(DFT) and CCSD(T) interaction
energies for linear MgNH configurations. Lines with
black squares refer to Mg-NH geometries, while empty squares
refer to Mg-HN geometries.} \label{saptcomp}
\end{figure}

Fig.\ \ref{figmgnh} shows the potential energy surface for
Mg--NH obtained from SAPT(DFT) calculations. There are minima
at both linear geometries and a saddle point between them, with
a barrier of only 24 cm$^{-1}$. The agreement between CCSD(T)
and SAPT(DFT) methods for linear configurations is very good
for the Mg-NH geometry, where SAPT(DFT) overestimates the
RCCSD(T) well depth by 6\%. The agreement is slightly worse for
the Mg-HN geometry, where SAPT(DFT) underestimates the well
depth by 9\% (see Fig.\ \ref{saptcomp}).

The ion-pair state does not reach negative energies until
distances $R<2.6$ \AA. The dispersion-bound state is strongly
repulsive at such distances. We have carried out multireference
self-consistent field (MCSCF) calculations of the two states of
$^3A''$ symmetry in the region of their avoided crossing for a
range of angles using a cc-pVQZ basis set \cite{Iron:2003,
Dunning:1989}. The lowest barrier for crossing onto the
ion-pair state occurs for an angle $\sim 110^\circ$ at a
distance of $R\sim2.4$ \AA\ and an energy of $+2200$ cm$^{-1}$
with respect to the atom-molecule threshold. The singlet
ion-pair state is about 700 cm$^{-1}$ shallower than the
triplet near its equilibrium geometry and will therefore cross
the dispersion-bound state at even higher energies.

The collision energies of importance to sympathetic cooling are
in the range between 1 $\mu$K and 100 mK (0.06 cm$^{-1}$). At
such energies we believe that the barrier separating the wells
of the ion-pair and dispersion-bound states is wide and high
enough to neglect the conversion from MgNH to Mg$^+$NH$^-$ and
to perform collision calculations only on the dispersion-bound
surface.

\section{Conclusions}

In this paper we have presented an overview of the interaction
potentials of alkali-metal and alkaline-earth atoms with NH
molecules in their ground $^3\Sigma^-$ state. The interaction
potentials of quartet states of AlkNH systems are strongly
anisotropic, with deep wells at Alk-NH geometries. The bonding in
the well region involves strong mixing of the $s$ and $p$ orbitals
of the alkali-metal atom and is thus covalent in nature. For
geometries close to Alk-NH configurations the quartet states are
crossed by ion-pair doublet states in the energetically accessible
region. Because of the presence of the ion-pair state, the lowest
doublet adiabatic potential energy surface has a very deep well. The
anisotropies for both doublet and quartet states are so large that
it is unlikely that sympathetic cooling of NH by alkali-metal atoms
will be successful for molecules in low-field-seeking states.

For alkaline-earth atoms the interaction potentials are much
shallower and less anisotropic, especially for BeNH and MgNH.
The $sp$ mixing is much weaker and the bonding is dominated by
induction and dispersion forces. For MgNH the anisotropy is
comparable to or smaller than the rotational constant of NH,
$B_e$. For CaNH and SrNH the anisotropy is rather larger, of
the order of $4B_e$ for CaNH and $11B_e$ for SrNH. The
relatively weak anisotropy raises the hope that alkaline earth
atoms could be used for sympathetic cooling of NH molecules.

The dispersion-bound $^{3}\Sigma^{-}$ states of AeNH systems
are crossed by singlet and triplet ion-pair states $^{1}\Pi$
and $^{3}\Pi$. At nonlinear geometries the $^{3}\Sigma^{-}$
state becomes $^3A''$ and there is a component of the $^3\Pi$
state of the same symmetry. The absolute minimum thus has
ion-pair character in all cases. For CaNH and SrNH the ion-pair
state crosses the dispersion-bound state in the energetically
accessible region, so that the adiabatic potential energy
surfaces have a single-minimum structure with a deep potential
well and strong anisotropy. For BeNH and MgNH, however, the
crossing occurs on the repulsive wall of the potential of the
dispersion-bound state and the deep ion-pair well is likely to
be inaccessible in low-energy collisions. We have calculated a
full 2-dimensional potential energy surface for MgNH and
verified that the crossing occurs on the repulsive wall at all
geometries.

The BeNH and MgNH systems are thus promising candidates for
sympathetic cooling. In future work we will carry out collision
calculations on MgNH to explore this further.

\section*{ACKNOWLEDGMENTS}
The authors are grateful to EPSRC and the Czech Science Foundation
(grant No.\ QUA/07/E007) for funding under the collaborative
projects QuDipMol and CoPoMol of the ESF EUROCORES Programme
EuroQUAM. PS also acknowledges support from the Ministry of
Education, Youth and Sports of the Czech Republic (Research project
no.\ 0021620835).


\end{document}